\documentclass[twocolumn]{aastex631}

\newcommand{\hi}{H\,{\sc i}}
\newcommand{\hii}{H\,{\sc ii}}
\newcommand{\oi}{O\,{\sc i}}
\newcommand{\oiii}{O\,{\sc iii}}
\newcommand{\cii}{C\,{\sc ii}}
\newcommand{\ciii}{C\,{\sc iii}]}
\newcommand{\src}{GS-z14}
\newcommand{\lya}{Ly$\alpha$}
\usepackage{xspace}
\let\oldAA\AA
\renewcommand{\AA}{\text{\oldAA}\xspace}

\begin{document}

\title{Dissecting the massive pristine, neutral gas reservoir of a remarkably bright galaxy at $z=14.179$}

\author[0000-0002-0786-7307]{Kasper~E.~Heintz}
\affiliation{Cosmic Dawn Center (DAWN), Denmark}
\affiliation{Niels Bohr Institute, University of Copenhagen, Jagtvej 128, 2200 Copenhagen N, Denmark}
\affiliation{Department of Astronomy, University of Geneva, Chemin Pegasi 51, 1290 Versoix, Switzerland}
\author[0009-0001-2808-4918]{Clara~Pollock}
\affiliation{Cosmic Dawn Center (DAWN), Denmark}
\affiliation{Niels Bohr Institute, University of Copenhagen, Jagtvej 128, 2200 Copenhagen N, Denmark}

\author[0000-0002-7595-121X]{Joris~Witstok}
\affiliation{Cosmic Dawn Center (DAWN), Denmark}
\affiliation{Niels Bohr Institute, University of Copenhagen, Jagtvej 128, 2200 Copenhagen N, Denmark}

\author[0000-0002-6719-380X]{Stefano~Carniani}
\affiliation{Scuola Normale Superiore, Piazza dei Cavalieri 7, I-56126 Pisa, Italy}

\author{Kevin~N.~Hainline}
\affiliation{Steward Observatory, University of Arizona, 933 North Cherry
Avenue, Tucson, AZ 85721, USA}

\author[0000-0003-2388-8172]{Francesco~D'Eugenio}
\affiliation{Kavli Institute for Cosmology, University of Cambridge, Madingley Road, Cambridge CB3 0HA, UK}
\affiliation{Cavendish Laboratory, University of Cambridge, 19 JJ Thomson Avenue, Cambridge CB3 0HE, UK}

\author[0009-0005-4175-4890]{Chamilla~Terp}
\affiliation{Niels Bohr Institute, University of Copenhagen, Jagtvej 128, 2200 Copenhagen N, Denmark}

\author{Aayush~Saxena}
\affiliation{Department of Physics, University of Oxford, Denys Wilkinson Building, Keble Road, Oxford OX1 3RH, UK}
\affiliation{Department of Physics and Astronomy, University College London, Gower Street, London WC1E 6BT, UK}

\author{Darach~Watson}
\affiliation{Cosmic Dawn Center (DAWN), Denmark}
\affiliation{Niels Bohr Institute, University of Copenhagen, Jagtvej 128, 2200 Copenhagen N, Denmark}

\begin{abstract}
At cosmic dawn, the first stars and galaxies are believed to form from and be deeply embedded in clouds of dense, pristine gas. Here we present a study of the JWST/NIRSpec data of the most distant, spectroscopically confirmed galaxy observed to date, JADES-GS-z14-0 (\src\ for short), at $z=14.179$, combined with recently reported far-infrared measurements of the [\oiii]-$88\mu$m and [\cii]-$158\mu$m line transitions and underlying dust-continuum emission. Based on the observed prominent damped Lyman-$\alpha$ (DLA) absorption profile, we determine a substantial neutral atomic hydrogen (\hi) column density, $\log (N_{\rm HI} / {\rm cm^{-2}}) = 22.27^{+0.08}_{-0.09}$, consistent with previous estimates though seemingly at odds with the dynamical and gas mass of the galaxy. Using various independent but complementary approaches, considering the implied neutral gas mass from the DLA measurement, the star-formation rate surface density, and the metal abundance, we demonstrate that the total gas mass of \src\ is of the order $\log (M_{\rm gas} / M_\odot) = 9.8\pm 0.3$. This implies a substantial gas mass fraction, $f_{\rm gas} \gtrsim 0.9$ and that the bulk of the interstellar medium (ISM) is in the form of \hi. 
We show that the derived gas mass is fully consistent with the non-detection of [\cii]-$158\mu$m, assuming an appropriate scaling to the neutral gas. The low dust-to-gas ratio, $A_V/N_{\rm HI} = (1.3\pm 0.6)\times 10^{-23}$\,mag\,cm$^2$, derived in the line-of-sight through the DLA further indicates that the absorbing gas is more pristine than the central, star-forming regions probed by the [\oiii]-$88\mu$m emission.
These results highlight the implications for far-infrared line-detection searchers attainable with ALMA and demonstrate that the bright, relatively massive galaxy \src\ at $z=14.179$ is deeply embedded in a substantial, pristine \hi\ gas reservoir dominating its baryonic matter content. 
\end{abstract}

\keywords{dark ages, reionization, first stars --- galaxies: formation, high-redshift}


\section{Introduction} \label{sec:intro}

The first stars and galaxies are believed to have formed within the first 150-250 Myr after the Big Bang, at redshifts around $z=15-20$ \citep[see e.g.,][for a review]{Robertson22}. This process is mainly driven by the inflow of pristine gas, mostly in the form of neutral atomic hydrogen (\hi), onto dark matter halos. With the advent of the {\em James Webb Space Telescope} (JWST), we are now starting to uncover and spectroscopically characterize a substantial population of galaxies close to this first assembly stage at $z=10-14$ \citep{CurtisLake23,Bunker23_gnz11,ArrabalHaro23,Castellano24,Hsiao24a,Hsiao24b,Witstok24,Carniani24a,Carniani24b,Zavala25}, and potentially even more distant sources at $z>15$ as indicated by their photometry \citep[e.g.,][]{Naidu22,Donnan23,Castellano22,Castellano23,Bouwens23,Atek23b,Harikane23,Austin24,Kokorev24,Whitler25}. 

The redshifts for these distant sources can be difficult to accurately pin-point with JWST near-infrared spectroscopy alone, as the Lyman-$\alpha$ break redshifts were found already in early JWST data to systematically overestimate the emission-line redshifts \citep[e.g.,][]{CurtisLake23,ArrabalHaro23,Fujimoto23_ceers,Finkelstein24}. This effect was later shown to originate from dense, neutral gas reservoirs in and around these galaxies, some with substantial \hi\ column densities, $N_{\rm HI} \gtrsim 10^{22}$\,cm$^{-2}$, in the form of extremely damped Ly$\alpha$ (DLA) absorption line profiles \citep[][]{Heintz24_DLA,Umeda23,DEugenio24,Hainline24b,Witstok24}. These deep DLA absorption profiles even systematically bias the photometrically-derived redshifts due to their high incidence rate in the overall $z \gtrsim 9$ galaxy population \citep[][]{Heintz25,Asada24}. These measurements indicate that we are starting to probe the primordial matter that drives the formation of these galaxies and fuels their intense, early star-formation activity.

Due to the intrinsic faintness of rest-frame UV emission lines covered by JWST/NIRSpec for galaxies at $z>10$, and with the added complication of potential strong DLA absorption, a promising alternative avenue of accurately determining the redshift of these sources is via bright far-infrared emission lines. While the first efforts with the Atacama Large Millimetre/sub-millimetre Array (ALMA) mostly returned non-detections \citep{Fujimoto23_alma}, the [\oiii]-$88\mu$m emission of the galaxy GHZ2 were detected at $z=12.33$ \citep[][]{Zavala24}. Further, ALMA observations of JADES-GS-z14-0 (\src\ for short), revealed a significant detection of the [\oiii]-$88\mu$m line \citep[][]{Schouws24,Carniani24b} lining up precisely with a marginal ($3.6\sigma$) NIRSpec detection of the \ciii $\lambda 1907, 1909 \, \AA$ doublet \citep{Carniani24a}. The line-redshift was thus confirmed to be at $z=14.1796\pm 0.0006$, substantially lower ($\Delta z = 0.14$) than the photometric and Ly$\alpha$ break redshifts due to the strong DLA recovered in the JWST spectrum \citep[][]{Carniani24b}. Deep ALMA follow-up observations targeting the far-infrared [\cii]-$158\mu$m line revealed a non-detection, however, which potentially indicate a surprisingly low gas fraction in \src\ \citep[][]{Schouws25} compared to other high-redshift sources \citep[e.g.,][]{Heintz23_JWSTALMA,Aravena24,Algera25}.

In this Letter, we reanalyze the JWST/NIRSpec Prism spectrum of \src\ in the context of its \hi\ gas content, corroborated by recent far-infrared line emission and continuum measurements. In Sect.~\ref{sec:obs} we detail the observational data and in Sect.~\ref{sec:res} we present the main analysis and results. In Sect.~\ref{sec:disc}, we will discuss and conclude on our work. Throughout the paper, we assume the concordance $\Lambda$CDM cosmological model with $H_0 = 67.4$\,km\,s$^{-1}$\,Mpc$^{-1}$, $\Omega_{\rm m} = 0.315$, and $\Omega_{\Lambda} = 0.685$ \citep{Planck18}. This is also used to infer cosmological parameters such as the luminosity distance to \src\ and age of the Universe with the {\tt Astropy} software package \cite{Astropy}.

\section{Observations} \label{sec:obs}

\src\ was first identified by \citet[][]{Hainline24a} with a photometric redshift $z_{\rm phot} = 14.51$, refined to $z_{\rm phot} = 14.39$ in the high-redshift galaxy population study by \citet{Robertson24}. Multi-band JWST/NIRCam imaging was obtained as part of the JWST Advanced Deep Extragalactic Survey \citep[JADES;][Prog. IDs~1180 and~1210]{Eisenstein23a}, the First Reionization Epoch Spectroscopically Complete Observations \citep[FRESCO;][Prog. ID~1895]{Oesch23}, and of the JADES Origins Field \citep[][Prog. ID~3215]{Eisenstein23b}. The data reduction is outlined in \citep{Rieke23,Eisenstein23a,Eisenstein23b}. The photometry implicitly used in this article (to scale the spectrum) was obtained by modeling the galaxy with a S\'ersic profile, using the \texttt{forcepho} tool (Johnson B., in prep.), as described in \citet{Carniani24a}. The distinctive advantage of \texttt{forcepho} is the ability to model the galaxy surface brightness before combining individual frames into a mosaic \citep[e.g.,][]{Baker25}. The fiducial \texttt{forcepho} model yields a half-light semi-major axis of $R_{\rm UV} = 0.26$~kpc and a S\'ersic index of 1. The gravitational lensing  magnification from a foreground nearby galaxy at $z=3.47$ is estimated to be only $\mu=1.17$ \citep{Carniani24a}, so the lensing correction of 8~per cent is smaller than the 10~per cent systematic uncertainties of typical size measurements. \src\ has further been detected in $7.7\mu$m imaging with JWST/MIRI \citep{Helton24}, implying the presence of strong nebular emission lines.

The JWST spectrum was first presented in \cite{Carniani24a}. The data were obtained using the NIRSpec \citep{Jakobsen22} Micro-Shutter Assembly \citep[MSA;][]{Ferruit22}, with three-shutter slitlets (Prog. ID~1287, PI: Luetzgendorf). Here we focus on the prism spectrum\footnote{Available at \url{https://doi.org/10.5281/zenodo.12578542}}, which covers wavelengths $\lambda = 0.6-5.3\mu$m with a spectral resolving power $\mathcal{R}=30\text{--}300$, although the most important spectral region in this article, the Ly$\alpha$ break at $\lambda \approx 2~\mu$m, has $\mathcal{R}\sim 60$. The prism observations used the NRSIRS2 readout mode to mitigate pink noise \citep{Moseley10,Rauscher17}, with 19 groups per integration and 2 integrations per exposure, with three nodded exposures for accurate background subtraction. This sequence (about 2.33~h total integration) was repeated four times, for a final on-source time of 9.3~h. Following \citet{Carniani24a}, we apply a flux correction factor to the reduced spectrum to match the normalization of the \texttt{forcepho} photometry data using a simple polynomial $F_{\rm corr} = F_{\rm red} \times (0.18\times \lambda(\mu{\rm m}) + 1)$. This effectively takes into account and corrects for potential slit losses and/or uncertainties in the overall flux calibration.

Throughout this work, we adopt the spectral and physical galaxy measurements reported in \citet{Carniani24b}. These were obtained using the \texttt{prospector} spectral-energy distribution (SED) modeling tool \citep{Johnson21}, taking into account the full multi-wavelength spectroscopic and imaging data obtained for \src. This yields a star-formation rate $SFR=14.5\pm 3~\mathrm{M_\odot \, yr^{-1}}$ and a stellar mass of $\log (M_\star/\mathrm{M_\odot}) = 8.29^{+0.09}_{-0.10}$, assuming a \citet{Chabrier03} initial mass function (IMF).

Dedicated far-infrared follow-up observations with ALMA were carried out to search for the [\oiii]-$88\mu$m (Prog. ID: 2023.A.00037.S, PI: Schouws) and the [\cii]-$158\mu$m (Prog. ID: 2024.A.00007.S, PI: Schouws) line transitions. [\oiii]-$88\mu$m is detected at $>6\sigma$ with a line luminosity $L_{\rm [OIII]} = (2.0\pm 0.5)\times 10^{8}\,L_\odot$ \citep{Schouws24,Carniani24b}. Deep ALMA observations targeting the [\cii]-$158\mu$m line, however, resulted in a non-detection at a limit $<6\times 10^{7}\,L_\odot$ \citep[$3\sigma$;][]{Schouws25}, though still consistent with the typical high-redshift galaxy population \citep{Harikane20,Carniani20}. 
Here we examine the constraints delivered from the far-infrared line transitions in context of the \hi\ column density and physical properties to probe the full baryonic matter budget of \src. 

\begin{figure*}
    \centering
    \includegraphics[width=17cm]{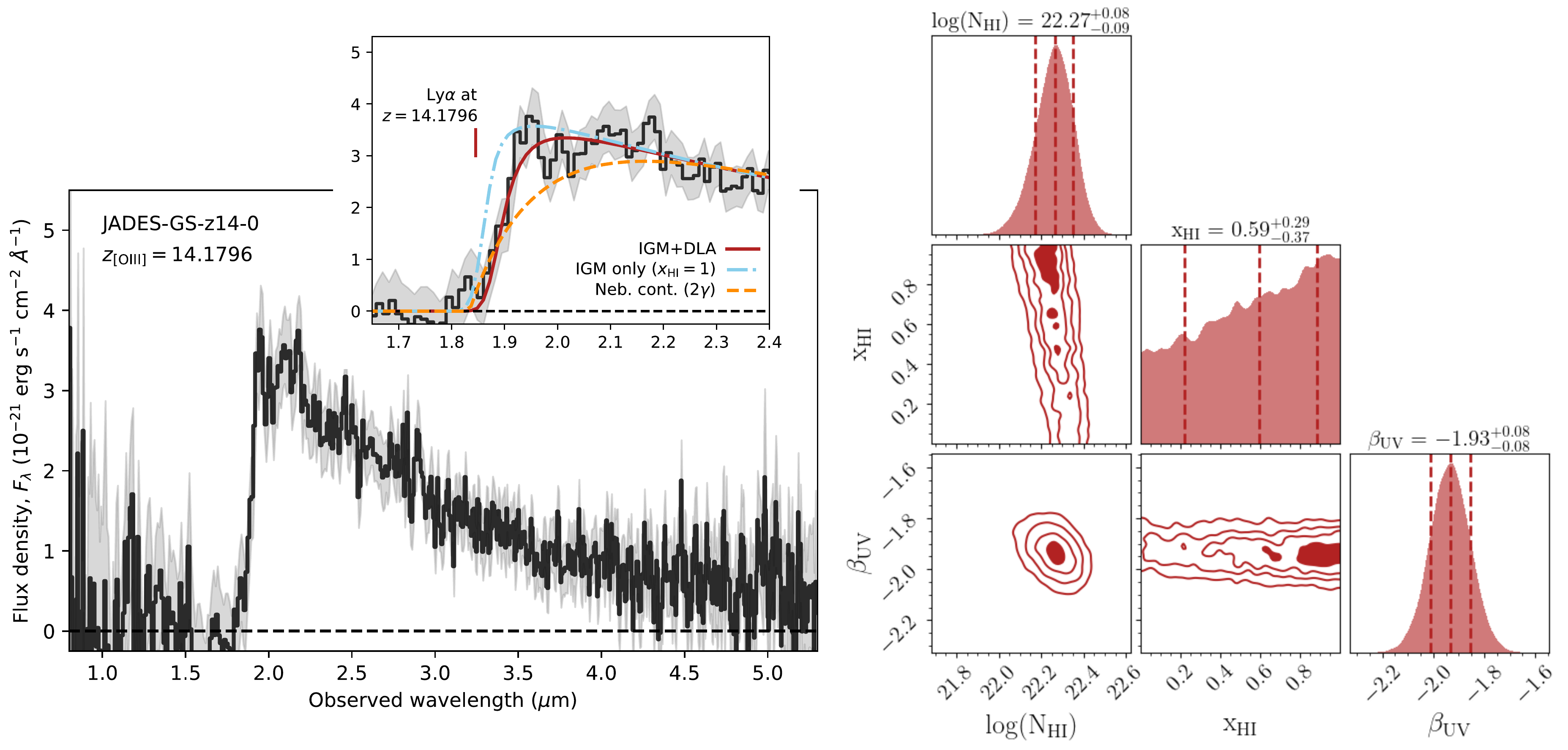}
    \caption{(Left): JWST/NIRSpec Prism 1D spectrum of \src. The photometrically-corrected flux density is shown by the black curve, and the associated uncertainty is shown by the grey-shaded region. In the inset is shown a zoom-in on the rest-frame UV part of the spectrum, with the best-fit DLA model overplotted (red solid curve), an IGM-only model (blue dot-dashed), and predictions for a $2\gamma$ nebular continuum emission (orange dashed). (Right): Cornerplot of the posterior distributions from the spectroscopic modeling, with the median and 16 to 84 percentiles marked.}
    \label{fig:spec}
\end{figure*}

\section{Analysis and results} \label{sec:res}

Based on the observed Ly$\alpha$ break, and given the lack of strong rest-frame UV/optical nebular emission lines, \citet{Carniani24a} inferred a redshift of $z_{\rm Ly\alpha,break} = 14.32^{+0.08}_{-0.20}$, with the large uncertainties due to the covariance between redshift and \hi\ column density. The significant detection of the far-infrared [\oiii]-$88\mu$m emission line reveals the systemic redshift to be $z=14.1796\pm 0.0006$ \citep{Carniani24b,Schouws24}. This measurement can break the degeneracy between redshift and \hi\ column density, and constrains that a substantial \hi\ column density arising in proximate neutral gas has imprinted a pronounced Ly$\alpha$ damping-wing absorption profile on the rest-frame UV continuum \citep{Carniani24b}. We model the DLA absorption with a Voigt profile following \citet[][see also Pollock et al. in prep.]{Heintz24_z54}, using the approximation from \citet{TepperGarcia06}, and convolve the best-fit model with the spectral resolution at each step during the optimization. The high \hi\ column densities of the absorbing gas effectively make this modeling only sensitive to the \hi\ column density, $N_{\rm HI}$, reflected in the broad, prominent \lya\ damping wing. Still, we include the unconstrained $x_{\rm HI}$ to fully capture the covariance between this and the other two parameters in the model. We assume that the rest-frame UV continuum can be approximated by a smooth power-law function, $F_\lambda \propto \lambda^{-\beta_{\rm UV}}$. Introducing more complex spectral shapes or prominent \lya\ emission does not significantly affect the derived $N_{\rm HI}$ \citep[][]{Heintz24_DLA}, at least at high \hi\ column densities \citep{Huberty25}. We also include a prescription of the neutral hydrogen fraction $x_{\rm HI}= n_{\rm HI}/n_{\rm H,tot}$ of the IGM on our modeling, which we expect to be fully neutral at $z\approx 14$ \citep[][]{MiraldaEscude98,Inoue14}. We use {\tt dynesty} nested sampling \citep{Speagle20} to sample the posteriors of the parameters $\log (N_{\rm HI})$, $x_{\rm HI}$, and $\beta_{UV}$, assuming a fixed DLA redshift of $z=14.1796$. 

The best-fit DLA model is shown in Fig.~\ref{fig:spec}, together with the posterior distributions on the derived quantities; $\log (N_{\rm HI} / {\rm cm^{-2}}) = 22.27^{+0.08}_{-0.09}$, $x_{\rm HI} > 0.41$ ($1\sigma$ from the $68\%$ highest density interval), and $\beta_{\rm UV} = -1.93\pm 0.08$.  
The derived $N_{\rm HI}$ is 0.3 dex higher than the column density inferred from the full spectro-photometric fitting by \citet{Carniani24a}, though still consistent within $3\sigma$. Setting the neutral hydrogen fraction to $x_{\rm HI} = 1$ as expected at $z=14$ does not significantly change the result since the DLA from the local \hi\ gas dominates the optical depth of the \lya\ absorption profile. Considering a model with only an IGM contribution ($x_{\rm HI}=1$) provides a statistically worse fit to the data with $\chi_\nu^2 = 1.69$, compared to $\chi_\nu^2 = 1.23$ for the fit with a DLA component. Additionally, the difference in Bayesian Information Criteria between the models $\Delta \rm BIC = 82$ indicates a strong statistical preference for the IGM+DLA model (see Fig.~\ref{fig:spec}). We also compare the DLA model to a purely nebular continuum model, dominated by the two-photon ($2\gamma$) emission at rest-frame UV wavelengths, assuming $T = 20,000$\,K and $n_e = 10^2$\,cm$^{-3}$ from \citet{Schirmer16}. This matches well the slope of the spectrum, but the strong UV turnover is inconsistent with the data ($\chi_\nu^2 = 1.53$).
With an absolute UV magnitude, $M_{\rm UV} = -20.81\pm 0.16$\,mag \citep[][]{Carniani24a}, \src\ is thus an exemplary case of a bright, star-forming galaxy embedded in a substantial neutral gas reservoir.

To gauge the total \hi\ gas mass of the galaxy, we use the measured \hi\ column density in the line of sight and assume that the half-mass radius of the neutral gas is $\sim 3\times$ the rest-frame UV size \citep[consistent with constraints of $z \sim 6$ galaxies from ALMA;][]{Fujimoto20,Fudamoto22}. For a spherically distributed gas, this yields an \hi\ gas mass of $M_{\rm HI} = 5\times 10^{9}\,M_\odot$, substantially higher than the previously reported upper bound inferred on the (molecular) gas mass, $M_{\rm mol} < 10^{9.2}\,M_\odot$ \citep[][]{Schouws25}. If we consider the metallicity-dependent relation converting the [\cii]-$158\mu$m line luminosity into a \hi\ gas mass \citep{Heintz21}, based on $\gamma$-ray burst sightlines through high-redshift star-forming galaxies, we derive an upper bound of $M_{\rm HI} < 10^{9.9}\,M_\odot$ ($3\sigma$). This suggests that the derived \hi\ gas masses are still consistent within the upper bound allowed by the non-detection of the [\cii]-$158\mu$m emission, taking into account the relatively low metallicity of the galaxy. However, this estimate of $M_{\rm HI}$ from the DLA $N_{\rm HI}$ is heavily dependent on the assumed geometry of the system. To corroborate the evidence for a substantial gas fraction in the system, we consider the Kennicutt-Schmidt relation \citep[][]{Kennicutt12}, universally connecting the SFR and gas surface density. We determine the star-formation rate surface density as $\Sigma_{\rm SFR} = {\rm SFR}_{\rm UV} / 2\pi R^2_{\rm UV}$, which given the half-light radius $R_{\rm UV} = 0.26$\,kpc reported by \citet{Carniani24a} yields $\log (\Sigma_{\rm SFR} / M_{\odot}\,{\rm yr}^{-1}\,{\rm kpc}^{-2}) = 1.5\pm 0.30$ (assuming an inflated $30\%$ error on the SFR). This implies a molecular gas surface density of $\log (\Sigma_{\rm H_2} / M_{\odot}\,{\rm pc}^{-2}) = 3.6\pm 0.3$ following the global-galaxy Kennicutt-Schmidt relation, which we convert to a molecular gas mass of $\log (M_{\rm H_2}/M_\odot) = 9.2\pm 0.3$ within the star-forming region, at the limit for the upper bound derived from the [\cii]-to-H$_2$ conversion \citep{Schouws25}. The global Kennicutt-Schmidt relation for the {\em total} (\hi\ + H$_2$) gas mass predicts a higher total gas surface density $\log (\Sigma_{\rm gas} / M_{\odot}\,{\rm pc}^{-2}) = 4.2\pm 0.3$ and total gas mass $\log (M_{\rm gas}/M_\odot) = 9.8\pm 0.3$ within the star-forming region.
Taken at face value, this implies a large gas fraction $f_{\rm gas} = M_{\rm gas} / (M_\star + M_{\rm gas}) \gtrsim 0.9$. Further, this estimate is consistent with the bulk of the interstellar medium (ISM) in \src\ being in the form of \hi\ ($M_{\rm HI} = M_{\rm gas}-M_{\rm H_2} = 5\times 10^{9}\,M_\odot$, $M_{\rm HI} \approx 3\times M_{\rm H_2}$) exactly as predicted from the DLA \hi\ column density. 

Alternatively, we can determine the metal mass of \src\ from the measured line detection of [\oiii]-$88\mu$m, and convert into a gas mass based on the metallicity quantified as an oxygen abundance, $\log ({\rm O/H}) = 7.92$ (for $Z/Z_\odot = 0.17$, and a solar oxygen abundance of $12+\log ({\rm O/H})_\odot = 8.69$; \citealt{Asplund09}). The metal mass in the ISM is defined as $M_Z = M_{\rm gas}\times Z/Z_\odot \times Z_{\rm \odot,ref}$ with $Z_{\rm \odot,ref} = 0.018$. Using the set of observations from the local {\em Herschel} dwarf galaxy survey \citep{Cormier15} and smoothed particle hydrodynamics simulations applied to cosmological zoom-in simulations \citep{Olsen17}, we find that the [\oiii]-to-$M_Z$ calibration is anti-correlated with the $L_{\rm [OIII]} / L_{\rm [CII]}$ line luminosity ratio \citep[see also][]{Heintz23_mz}, such that $\log (M_Z / L_{\rm [OIII]}) = -1.0\pm 0.2$ for $L_{\rm [OIII]} / L_{\rm [CII]}>3$. This yields a total metal mass in the ISM of $M_Z = (2.0\pm 0.6)\times 10^{7}\,M_\odot$, which implies a total gas mass of $\sim 10^{9.8}\,M_\odot$ when taking the metallicity of \src\ into account, consistent with the above estimates. 

\begin{figure}
    \centering
    \includegraphics[width=8.8cm]{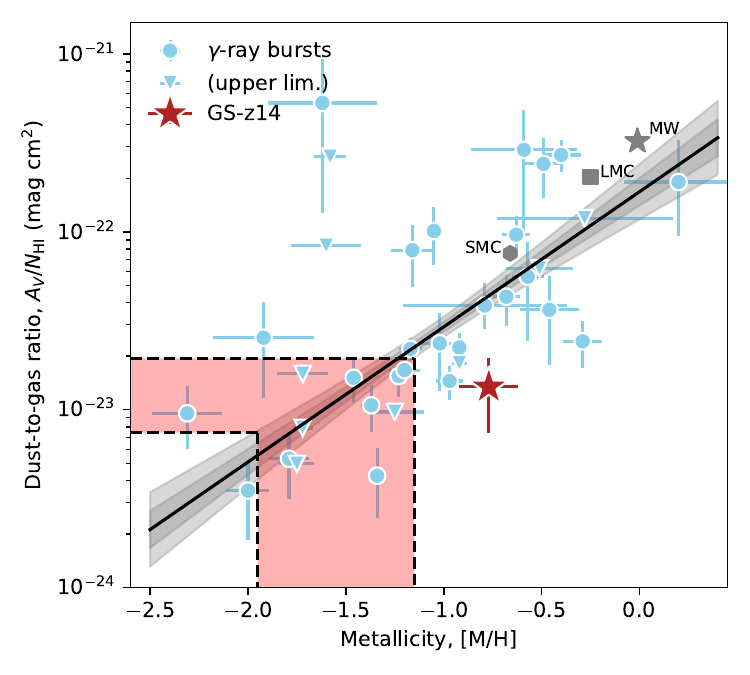}
    \caption{Dust-to-gas ratio, $A_V/N_{\rm HI}$, as a function of gas-phase metallicity, [M/H]. \src\ is shown by the red star symbol, and the high-redshift $\gamma$-ray burst sightlines through star-forming galaxies at $z=1.7-6.3$ as the blue circles (measurements) or triangles (upper limits). The best-fit relation for the $\gamma$-ray burst sample is shown by the black solid line, with the $1\sigma$ and $2\sigma$ uncertainty indicated by the dark- and light-grey shaded region. The red shaded region marks the expected gas-phase metallicity from the relation, assuming the measured dust-to-gas ratio of \src, $A_V/N_{\rm HI} = (1.3\pm 0.6)\times 10^{-23}$\,mag cm$^2$.}
    \label{fig:dtg}
\end{figure}

To investigate in more detail the composition of the absorbing gas, we determine the dust-to-gas ratio, $A_V/N_{\rm HI}$. Here, $N_{\rm HI}$ is derived from the DLA model, and the visual extinction $A_V$ is determined from the spectral slope. Assuming a steep, intrinsic power-law with $\beta_{\rm UV} = -3$ requires $A_V = 0.3$\,mag for a SMC-like reddening curve \citep{Gordon03} to match the observed rest-frame UV spectral slope. This quantity is effectively the maximum allowed amount of dust in the line-of-sight that will extinguish the light, since the intrinsic spectral slope will be redder for an slightly older stellar population on average. This upper bound is also consistent with the visual attenuation $A_V = 0.25\pm 0.10$ inferred from the full spectro-photometric modeling of the SED and line emission for \src\ \citep[][which we assume in the following]{Carniani24b}. This yields a dust-to-gas ratio of $A_V/N_{\rm HI} = (1.3\pm 0.6)\times 10^{-23}$\,mag\,cm$^{2}$. We compare this to equivalent measures from the compilation of $\gamma$-ray burst sightlines through high-redshift ($z=1.7-6.3$) star-forming galaxies presented in \citet{Heintz23_GRB} and to average value observed in the Milky Way and the Small and Large Magellanic Clouds \citep[SMC and LMC;][]{Konstantopoulou24} in Fig.~\ref{fig:dtg}. It is evident that \src\ has a relatively low dust-to-gas ratio, consistent with the most metal-poor ([M/H]$<-1.0$) $\gamma$-ray burst sightlines. This suggest that this particular sightline either has a low dust-to-gas ratio due to inefficient dust production (or more violent dust destruction mechanisms) or that the DLA mainly trace more pristine gas than the galaxy itself. The empirical, best-fit relation indicate a gas-phase metallicity of $-1.95 < {\rm [M/H]} < -1.15$ (or $12+\log({\rm O/H}) = 6.75-7.55$). The inferred lower gas-phase metallicity from this approach compared to the emission-line metallicity is similar in methodology and consistent with the results from \citet{DEugenio24} for GS-z12 at $z=12.5$. We will discuss the observed dust deficit further in Sect.~\ref{sec:disc} below. 

\begin{figure*}
    \begin{minipage}[c]{0.65\textwidth}
    \includegraphics[width=\textwidth]{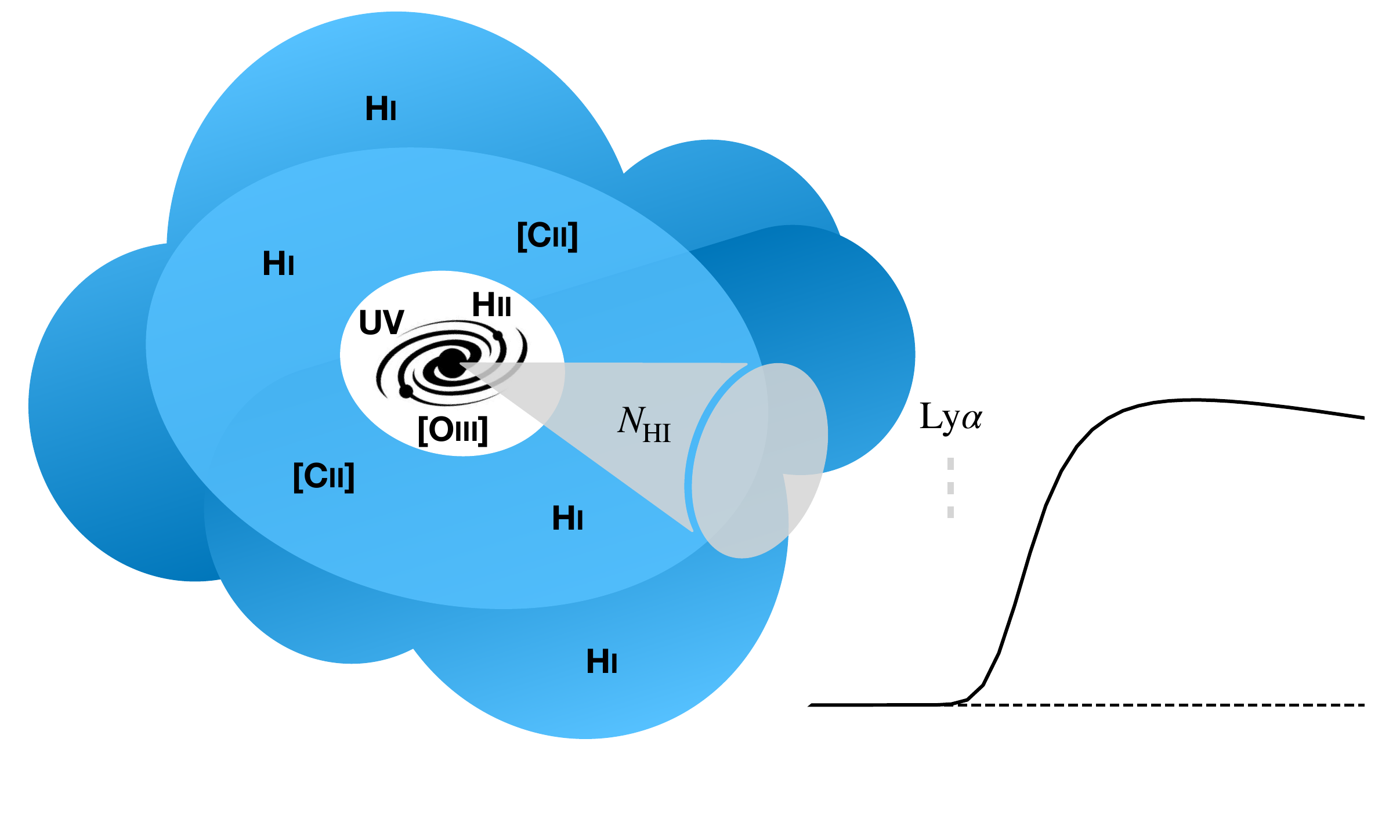}
  \end{minipage}\hfill
  \begin{minipage}[c]{0.3\textwidth}
    \caption{Schematic of the gas components seen in absorption and emission for \src. The central star-forming \hii\ regions also emit most of the rest-frame UV and [\oiii] emission. The more extended neutral gas region likely dominates the observed \hi\ column density, and more diffuse [\cii] emission due to the lower metallicity of the gas. The compact size of the young stellar population increases the ionization parameter $U$ and the [\oiii]/[\cii] line ratio. Inspired by \citet{Harikane20}.} 
    \label{fig:gal}
  \end{minipage}
\end{figure*}

\section{Discussion and conclusions} \label{sec:disc}

We have examined in detail the gas content of the most distant, spectroscopically confirmed galaxy observed to date, \src\ at $z=14.179$ \citep{Carniani24a,Carniani24b,Schouws24}. This galaxy is relatively bright, with an absolute UV magnitude $M_{\rm UV} = -20.8$\,mag, a stellar mass $M_\star \sim 10^{8.3}\,M_\odot$, and shows evidence for chemically enriched gas ($Z/Z_\odot \sim 0.17$). We determined the total \hi\ column density for \src\ to be $\log (N_{\rm HI}/{\rm cm^{2}})=22.27^{+0.08}_{-0.09}$, integrated over the line-of-sight. Assuming a simple spherical geometry, this implies a substantial \hi\ gas mass of $M_{\rm HI} = 5\times 10^{9}\,M_\odot$.
While a spherical geometry is an over-simplification, the high detection rate of DLA around galaxies at $z>10$ \citep{Heintz25} almost certainly implies a high covering factor, closer to a spherical distribution than to a thin disc. Importantly, we found this large \hi\ gas mass estimate to be consistent with the expected total gas mass from the star-formation rate surface density and the universal Kennicutt-Schmidt relation and within the limits allowed by the non-detection of the [\cii]-$158\mu$m line emission \citep{Schouws25}. These independent but complementary measurements all pointed to a total gas mass of the order $\log (M_{\rm gas}/M_\odot) = 9.5\pm 0.3$, implying a substantial gas mass fraction, $f_{\rm gas}\sim 0.7-0.9$. Assuming a bottom-light
or top-heavy IMF \citep[e.g.,][]{Bate23,Bate25,Katz25} would increase the fraction even further.
This implies that the bulk ISM, and the baryonic matter content of \src\ overall, is largely dominated by \hi. Based on the low visual attenuation, $A_V = 0.25\pm 0.10$, we determined that the metallicity of the DLA is lower than that of the central star-forming region. This suggested that \src\ is embedded in a substantial, pristine neutral gas reservoir. 

Previous estimates reporting a low gas fraction, $f_{\rm gas} \lesssim 0.7$, for \src\ are based on the dynamical mass estimated within the [\oiii]-$88\mu$m emitting region ($\log (M_{\rm dyn}/M_\odot) = 9.0\pm 0.2$) and the non-detection of the [\cii]-$158\mu$m emission line \citep{Carniani24b,Schouws25}. We argue that since [\oiii]-$88\mu$m solely originates from the central, most intense star-forming regions (as is also the case for \ciii\,$\lambda 1909$), this line transition is arguably not a reliable tracer of the total baryonic matter in the galaxy. Indeed, the rest-frame UV and the [\oiii]-$88\mu$m emission has been observed to be much more spatially compact than the [\cii]-$158\mu$m emitting region \citep[e.g.,][]{Harikane20,Fujimoto20,HerreraCamus21,Fudamoto22}. The fact that it is consistent with the inferred stellar mass from the SED modeling \citep{Carniani24b} instead supports this line mostly tracing the central star-forming component. The upper bound on the molecular gas mass inferred from the non-detection of [\cii]-$158\mu$m \citep{Schouws25} will also likely underestimate the total gas mass of the galaxy, given that [\cii] predominantly traces the neutral atomic gas, in particular in low-metallicity regions \citep{Madden93,Heintz22}. This is because the ionization potential of neutral carbon (11.26 eV) is below that of neutral hydrogen (13.6 eV), such that any inherent ISM radiation field will ionize carbon to C$^+$ in the neutral ISM. Indeed, the substantial \hi\ column density suggest that there is a large abundance of neutral H outside the central \hii\ region. We show a schematic of this proposed interpretation in Fig.~\ref{fig:gal}. 
We note that the H$_2$ fraction may be suppressed in these galaxies by the lack of dust compared to lower-redshift galaxies, and that the lack of [\cii] could also be explained by a low C/O abundance ratio \citep{Carniani24b}, which is expected for low-metallicity galaxies \citep[e.g.,][]{Nicholls17,TJones23,ArellanoCordova22,Curti24z94}, or the high ionization parameter $\log U = -2.4$ which suppresses [\cii] in favor of \ciii.


The analytical model developed for \src\ by \citet{Ferrara24} of its star formation history and general physical properties is generally consistent with the observed quantities \citep{Carniani24b}, and particularly predicts a low dust-to-gas mass ratio of $10^{-3.5}$. This is exactly at the limit we predict from the total gas mass and the upper bound on the dust mass $<10^{6}\,M_\odot$ from the far-infrared continuum \citep{Schouws24}. We note that the observed dust-to-gas mass ratio ($<10^{-3.5}$) is $\sim 10\times$ lower than the Galactic average \citep{Konstantopoulou24}, but consistent with the general trend seen at the local to high-redshift universe considering its low metallicity \citep{Heintz23_GRB}. Further, the line-of-sight dust-to-gas ratio, $A_V/N_{\rm HI}$, is equally lower than the Galactic average at $4.5\times 10^{-22}$\,mag\,cm$^2$ \citep{Watson11}. These independent measures thus validate the low dust content of \src, but suggest that it is primarily driven by the metallicity of the galaxy.  

Our results have important implications for the follow-up strategies with both JWST and ALMA for characterizing in depth the gas and metal abundances of very high-redshift ($z>10$) galaxies. First, the typical lower metallicities of the galaxies affect the C/O abundance ratio and how in particular the far-infrared [\cii]-$158\mu$m emission scales with the neutral gas mass \citep{Katz22}, while the [\oiii]-$88\mu$m transition remains a valid tracer of the star-formation activity of even the most distant galaxies. This greatly motivates targeting the [\oiii]-$88\mu$m line transition \citep[e.g.][]{Bouwens22} to accurately pin-point the redshift of very high-redshift galaxies. We caution, however, that the gas mass inferred from the dynamics of [\oiii] will likely underestimate the true total baryonic matter content of the sources. Second, the typical strong DLAs detected in the spectra of galaxies at $z>10$ \citep[][]{Heintz24_DLA,Heintz25,Umeda23,DEugenio24,Hainline24b,Witstok24,Asada24} need to be taken into account when designing the far-infrared line scan configuration, which will be at a lower redshift than implied from the \lya\ break alone. Finally, quantifying the metal abundance in the absorbing gas traced by the DLA will help disentangle whether this is mainly pristine, intergalactic gas or consistent with the chemically enriched interstellar inferred from the SED and nebular emission line ratios. For the substantial \hi\ column densities $N_{\rm HI}\gtrsim 10^{22}$\,cm$^{-2}$ and metallicities $Z/Z_\odot \sim 10\%$ seen in \src\ and other spectroscopically confirmed galaxies at $z>10$, we would expect to be able to detect the low-ion metal absorption lines from \oi\,$\lambda 1302$ or \cii\,$\lambda 1334$, tracing the metal abundance of the neutral gas, in the higher resolution JWST/NIRSpec grating spectra. The main challenge will be to achieve high enough signal-to-noise in the rest-frame UV continuum to robustly measure the optical depth and thereby the column densities of the metal absorption lines, which requires substantial exposure times even for the intrinsically bright \src\ due to the large cosmological distances. This measure will, however, definitely establish whether the massive DLAs seen in most high-redshift spectra are predominantly tracing primordial or chemically enriched neutral atomic gas at cosmic dawn. 




\begin{acknowledgments}
\footnotesize
We would like to thank Johan Fynbo and Peter Laursen for insightful discussions about the interpretation of the results presented in this work. We also acknowledge and appreciate the substantial amount of work from the JADES collaboration in defining and carrying out the JWST/NIRSpec spectroscopic follow-up of the most distant galaxies uncovered to date. This work has received funding from the Swiss State Secretariat for Education, Research and Innovation (SERI) under contract number MB22.00072. The Cosmic Dawn Center (DAWN) is funded by the Danish National Research Foundation under grant DNRF140. FDE acknowledges the ERC Advanced Grant 695671 ``QUENCH'' and support by the Science and Technology Facilities Council (STFC) and by the UKRI Frontier Research grant RISEandFALL. 

This work is based on observations made with the NASA/ESA/CSA James Webb Space Telescope. The data were obtained from the Mikulski Archive for Space Telescopes at the Space Telescope Science Institute, which is operated by the Association of Universities for Research in Astronomy, Inc., under NASA contract NAS 5-03127 for JWST. These observations are associated with programs \#1210, \#1287, and \#3215. This work was furthermore based on observations taken by the Atacama Large Millimeter/submillimeter Array (ALMA). ALMA is a partnership of ESO (representing its member states), NSF (USA) and NINS (Japan), together with NRC (Canada), MOST and ASIAA (Taiwan), and KASI (Republic of Korea), in cooperation with the Republic of Chile. The Joint ALMA Observatory is operated by ESO, AUI/NRAO and NAOJ.
\end{acknowledgments}

%

\vspace{5mm}
\facilities{JWST(NIRSpec), ALMA}


\software{Astropy \citep{Astropy},  
          Matplotlib \citep{Matplotlib},
          Numpy \citep{Numpy}}





\bibliography{ref}{}
\bibliographystyle{aasjournal}



\end{document}